\documentclass{llncs}
\usepackage{graphicx}
\usepackage{programs}
\usepackage{paralist}
\usepackage{amssymb}
\usepackage{booktabs}
\usepackage{multibib}
\usepackage{harvard}
\newcites{sec}{References}
\usepackage{setspace}
\begin{document}
\title{Metamodel Instance Generation: \\ 
a systematic literature review}
\author{Hao Wu \and Rosemary Monahan \and James F. Power}
\author{Hao Wu\thanks{This work is supported by a John \& Pat Hume Scholarship from NUI Maynooth.} \and Rosemary Monahan \and James F. Power}
\institute{Department of Computer Science, National University of Ireland, Maynooth\\
\email{\{haowu,rosemary,jpower\}@cs.nuim.ie}}
\maketitle

\pagestyle{plain}
\thispagestyle{plain}
\begin{abstract}
Modelling and thus metamodelling have become increasingly important in Software Engineering through the use of Model Driven Engineering.  In this paper we present a systematic literature review of instance generation techniques for metamodels, i.e. the process of automatically generating models from a given metamodel.  We start by presenting a set of research questions that our review is intended to answer.  We then identify the main topics that are related to
metamodel instance generation techniques, and use these to initiate our literature search.  This search resulted in the identification of 34 key papers in the area, and each of these is reviewed here and discussed in detail.  The outcome is that we are able to identify a knowledge gap in this field, and we offer suggestions as to some potential directions for future research.
\end{abstract}
\section{Introduction}
Models are key representations in model-driven engineering (MDE), where software engineers use them to represent a system at an abstract level.  To a software engineer, interacting and refining the design of a system at an abstract level alleviates the complexity of a system from a code level.  The actual implementation can be manually or automatically generated from well designed models.  The introduction of a higher abstraction technique to MDE, metamodeling, allows software engineers to describe and refine their designs at a higher level.  To a software engineer, a \emph{model} can be regarded as an abstract representation of an aspect of a system, whereas a \emph{metamodel} describes a higher abstraction of meaning: it is a model that captures properties of a model.

\begin{figure}
\centering
\includegraphics[scale=0.4]{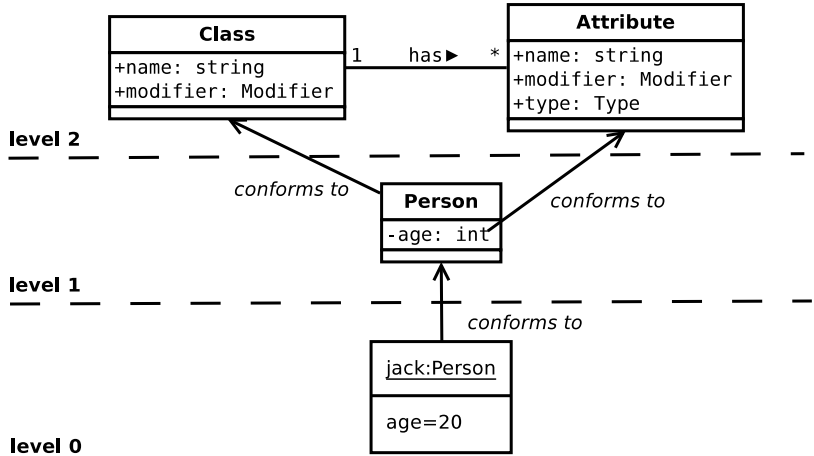}
\caption{An example of a model conforms to its metamodel}
\label{fig:example}
\end{figure}

An example of a model and metamodel can be seen in Figure \ref{fig:example}.  In Figure \ref{fig:example}, a person named ``jack" with an age of $20$ conforms to a model called $Person$.  The model $Person$ conforms to a model (metamodel) which sits at a higher level, and this metamodel has two metaclasses $Class$ and $Attribute$ (field $age$ conforms to $Attribute$).  The conformance here means that each element in the $Person$ model is a type of its metamodel.  For example, $Person$ is a type of $Class$, and $age$ is a type of $Attribute$.

In MDE, designing a large system begins with a variety of model designs, and thus revealing a system design fault at its model representation has become crucial for producing its implementation.  To show the correctness of a model, a large number of test cases will be needed.  These test cases can be manually or automatically produced based on different techniques.  In this paper, we focus on reviewing techniques that are used to generate test cases from a metamodel.  Thus, we limit our discussion of these techniques to those used to produce models at level 1 as depicted in Figure \ref{fig:example}.  

This paper is organised as follows.  We present the background of metamodeling and our research questions in Section 2.  In Section 3, we describe the methods for collecting and assessing the papers that we have studied, and present our findings.  In Section 4, selected papers are explained with reference to our research questions.  Additionally, we will identify the knowledge gap between model instance generation and metamodelling technique.  Finally, we conclude our literature review in Section 5.

\section{Background}
Metamodeling provides many advantages for software developers such as ease of use and a greater degree of abstraction.  For example, a metamodel can capture the abstract syntax of a programming language.  With these advantages, this technique has been widely adopted nowadays in software development.  However, the technique has one big disadvantage: it does not offer a way to automatically produce test cases for a metamodel.  Such test cases are models, and also referred to as $instances$ of a metamodel.  Thus, $Person$ in Figure \ref{fig:example} is an $instance$ of its metamodel, and producing test cases for a metamodel is equivalent to generating $instances$ ($models$) from a metamodel.

Furthermore, software engineers may add constraints on a metamodel to meet specific requirements, and this makes the disadvantage even greater since the models produced should not violate the constraints.  To overcome this limitation, researchers have proposed a variety of approaches and techniques.  The purpose of this literature review is to identify and analyse the strengths and weaknesses of these techniques.

To help us reduce the search scope of our literature review, we consider a metamodel in two dimensions: its structure representations and applications.  

In the structure dimension, we focus on what the various representations of a metamodel could be.  For example, a metamodel can be represented as a UML class digram, or viewed as an abstract syntax tree for a programming language \cite{umlinfra:2011,omg:mof:10} \cite{hei:09:gcm}.  To conform to the Meta-Object Facility (MOF) standard, we consider all metamodels in this paper as being represented as UML class diagrams \cite{omg:mof:10}.  Other kinds of models which are out of the scope of our work include state machines and sequence diagrams \cite{pilskalnsa07:ist,trong06:issre,nayak:09:mtc,briand10:stvr,mahdian09:jsm,sam:08:tssd}.

Metamodels we discuss in this paper must posses the features that a UML class diagram can have.  Thus, a metamodel must include one or many of the following features:

\begin{itemize}
\item[I. Classes]
\item[II. Attributes]
\item[III. Relationships (Generalisations and Associations) between classes]
\end{itemize}

As an example, the metamodel in Figure \ref{fig:example} depicts two classes ($Class$ and $Attributes$) and an association relationship $has$ indicates each $class$ can have multiple $attributes$.

In the application dimension, we focus on the area of model transformation testing because it requires a large number of sample instance as test cases, thus metamodel instance generation can be applied to test correctness of a model transformation.

Within these two dimensions, we will identify and discuss the underlying frameworks and algorithms that are used for generating metamodel instances.  We are also interested in the criteria for selecting models because measuring test cases is a crucial step when determining the qualities of a test suite.  Thus, based on these aspects of concern, we formed the following research questions.

\begin{enumerate}
\item[Research Question 1.]
	What are the main areas related to metamodel instance generation?
\item[Research Question 2.]
	What algorithms and theoretical frameworks are used for metamodel instance generation?
\item[Research Question 3.]
	What criteria are used for selecting instances?
\item[Research Question 4.]
	What tools exist to produce metamodel instances?
\end{enumerate}

\begin{figure}
\centering
\includegraphics[width=\textwidth]{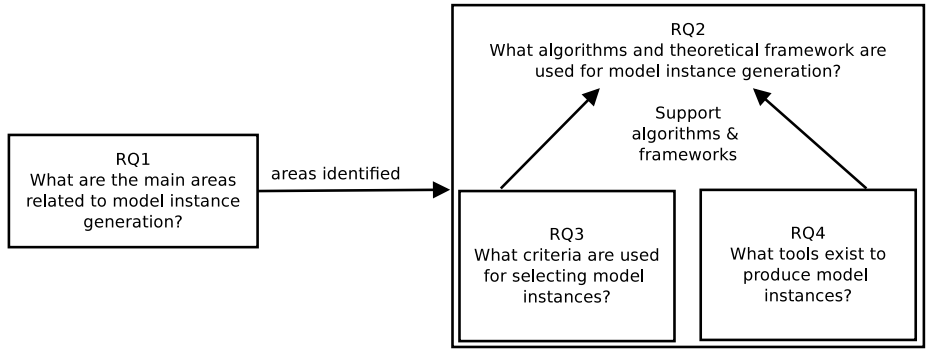}
\caption{The relationship between our Research Questions.}
\label{fig:rq}
\end{figure}

Figure \ref{fig:rq} shows the relationship between our four research questions.  First, by identifying the main areas that are related to metamodel instance generation, we gather and review the papers that describe algorithms and theoretical frameworks from these areas.  This includes reviewing the test criteria used and the tools that support the algorithms and theoretical frameworks.  Therefore, questions 3 and 4 are within the scope of question 2.

\section{Review Research Methods}
In this section, we describe the methodologies that we used for searching and assessing the papers we collected from different areas that are related to metamodel instance generation, and present selected papers.

In accordance with systematic review guidelines \cite{kitch:ppsr:04}, we have taken the following steps:
\begin{compactenum}
\item[1.] Identify the need for a systematic literature review
\item[2.] Formulate the review research questions
\item[3.] Conduct a comprehensive, exhaustive search for the primary studies
\item[4.] Assess the quality of the included studies
\item[5.] Extract the data from each included study
\item[6.] Summarise and synthesise the results of the study
\item[7.] Interpret the results and determine their applicability
\item[8.] Write-up the study as a report
\end{compactenum}

\subsection{Search terms}
Since metamodel instance generation is not a standalone area and overlaps many areas, our first research question which is ``What are the main areas related to model instance generation and how they are related?'' Although we defined our search scope according to various representation of a metamodel, some key papers that come from different areas can easily be overlooked.  In order to widen our search scope to include more areas, we form the following search terms which we conjoin to form our search.

\begin{enumerate}
\item (metamodel test case OR data generation OR metamodel instance generating OR generation OR model transformation testing OR model measurement)
\item (input test models OR automatic model generation OR generating)
\item (grammar testing OR grammar instance generation OR grammar sentence generation OR attribute grammar testing) 
\item (test data OR case generation from UML class diagram)
\item (XML schema testing)
\item (graph grammar instance OR graph grammar metamodel OR graph grammar testing)
\item (SAT/SMT OR constraint programming OR SAT/SMT UML/OCL verification)
\end{enumerate}

The search terms are derived from three angles.  First, the terms are directly derived from research questions, for example, keywords such as ``metamodel instance generation".  Second, others are formed based on various representations of a metamodel.  For example, ``metamodel represents a programming language" conveys information of an abstract syntax for a language.  Thus in this sense, it is very close to the area of grammar testing.  Third, the rest of the search terms are constructed to focus on how to deal with the constraints defined on a metamodel, for example SAT/SMT and constraint programming based approaches are particularly good at solving constraint problems.

The search process can be divided into two phases.  In the first phase, we group and input the search terms into search engines provided by the following databases:
\begin{compactenum}
\item[1.] Google scholar
\item[2.] ACM Digital library
\item[3.] IEEE Explore
\item[4.] ScienceDirect
\end{compactenum}
We reduce the number of papers by looking at the title and abstract of each paper.  In the second phase of the search, we review the citations from the results we get in the first phase for any relevant articles, and we repeat this process until no new papers are identified.  In addition, in order not to overlook any papers, we also search through possibly related conferences for a specific time period.  For each conference, we widen the time period from the first proceedings of that conference held to the latest proceedings.  The list of the conferences from a specific time period can be seen in Table \ref{tab:conference}.
\begin{table}
\begin{tabular}{ p{8cm} | l }
\toprule
Name of the Conferences  & Time Period Searched\\ \hline
Model Driven Engineering Languages and Systems & 2005 - 2012 \\ \hline
International Conference on Graph Transformation & 2004 - 2012 \\ \hline
International Conference on Fundamental Approaches to Software Engineering & 1998 - 2012\\ \hline
International Conference on Automated Software Engineering & 1997 - 2012\\ \hline
International Conference on Tools and Algorithms for the Construction and Analysis of Systems & 1995 - 2012\\ \hline
International Conference on Formal Methods & 1999 - 2011\\ \hline
International Conference on Software Testing Verification and Validation & 2008 - 2012\\ \hline
International Conference on Theory and Practice of Model Transformations & 2008 - 2012\\ \hline
International Conference on Tests and Proofs & 2007 - 2012\\ \hline
International Conference on Tools and Algorithms for the Construction and Analysis of Systems & 1995 - 2012\\
\bottomrule
\end{tabular}
\caption{Time period searched for each conference}
\label{tab:conference}
\end{table}

\subsection{Paper Selection Study}
A total of 23 papers collected from search results are selected for studying by judging their titles and abstracts.  However, 7 of them cannot be judged by doing so.  We choose to look at the contents of these papers in more detail, particularly, the examples used in these papers.  By studying the examples used in each paper, we identified the theme from each selected paper and recorded it.  We then performed a descriptive synthesis and recorded the data in a separate form for each study.  Thus, by synthesising the results from our search terms and selection study, we have gathered a total of 34 papers.

\subsection{Selected papers}
We review each paper for several times and a table is used for each paper to record main ideas of the paper.  A categorisation of the 34 papers that are found to be related to instance generation techniques across different areas is given in Table \ref{tab:paper}.  The areas included in this literature review are explained in Section 5.
\begin{table}
\begin{tabular}{  p{8cm} | p{3cm} }
	\toprule
	References & Area\\ \hline
	\citeasnoun{alan:arbc:03}	& Compiler testing \\ \hline
	\citeasnoun{fle:vme:04}, \citeasnoun{bro:06:mtg}, \citeasnoun{fle:qit:09}, \citeasnoun{lami:tat:07}, 	\citeasnoun{Sen:MMS:06}	& Model transformation testing \\ \hline
	\citeasnoun{hoff:11:gcm}, \citeasnoun{ehri:gimm:09}, \citeasnoun{win:tro:08}, \citeasnoun{heck:act:05} & Graph grammar \\ \hline	
	\citeasnoun{dan:02:alloy}, \citeasnoun{torlak:06:thedesign}, \citeasnoun{torlak:krm:07}, \citeasnoun{Ana:10:sos}, \citeasnoun{bord:ates:05}, \citeasnoun{Anas:uml2alloy:07}, \citeasnoun{sen:ocm:08}, \citeasnoun{sen:amgsmtt:09},
	\citeasnoun{Soe:11:TAD}, \citeasnoun{kuh:11:usekodkod}, \citeasnoun{mcquillan08:stvv}, \citeasnoun{soe:10:vumubs}, \citeasnoun{Soeken:11:EOD}, \citeasnoun{eth:11:Formula} & SAT/SMT based approaches\\ \hline
	\citeasnoun{cab:vuc:08}, \citeasnoun{cab:09:vuoc}, \citeasnoun{cab:uml2csp:07}, \citeasnoun{gon:emf2csp:12} & Constraint Programming approaches (CSP)\\ \hline
	\citeasnoun{bert:atd:07}, \citeasnoun{bert:sgx:06} & XML \\ \hline
	\citeasnoun{gogo:vuomu:05}, \citeasnoun{mou:urg:09}, \citeasnoun{lukman:agmtfmd:10}, \citeasnoun{elg:rrv:11} & Other \\ \bottomrule
\end{tabular}
\caption{Papers are grouped by 7 areas: compiler testing, model transformation testing, graph grammar, SAT/SMT based approaches, Constraint Programming, XML and other.}
\label{tab:paper}
\end{table}

\section{Discussion}
In this section, we discuss the papers that we collected from different areas.  This is done by answering the research questions defined in Section 2.  We also discuss main advantages and disadvantages of the different techniques at the end of this section.  

\subsection{RQ1. What are the main areas related to metamodel instance generation?}
In this literature review, we are interested in automatically generating instances (models) from a metamodel.  The generated instances are critical for reasoning about a metamodel.  In other words, they are mainly used for verifying the correctness of the system design.  The following paragraphs identify some of the main areas that are related to the topic of instance generation.

\subsubsection{1. Compiler testing}
A computer program can be syntactically parsed into a parse tree, and a parse tree is a graphic view of an instance of a programming language grammar.  In other words, a syntactically correct computer program must conform to its language's grammar.  This is similar to the concept of a sample model instance that conforms to a metamodel.  Thus, producing a sample model from its metamodel is related to the problem of generating programs from a given grammar, which is in the area of compiler testing \cite{bouj:ctcg:97}. 

\subsubsection{2. Model transformation testing}
Another area that is related to instance generation is model transformation testing.  Model transformations are 
are the essential feature in the approach of Model-Driven Engineering (MDE).  The process of model transformation takes as input a model which conforms to a source metamodel and transforms it to another model which conforms to a different target metamodel.  There are two ways to test this process: One way is testing the model transformation program code itself \cite{kuster:vmt:06}.  For example, making sure that all the statements are executed at least once.  This is a white-box testing technique and it requires testers to have knowledge of the internal logic or the code structure of the programs.  The other way is to produce a large set of sample models and run the transformation program with them to check the correctness of the result, regardless of internal structure of the transformation program.  This is a black-box testing technique since the testers do not have the knowledge about the code at all \cite{Bei:bt:1995}.  Therefore, the research work that has been done to automatically produce sample models in this area is closely related to our own interests.

\subsubsection{3. Graph Grammar}
Another area we are interested in is graph grammars (GG).  Graph grammars provide a theoretical foundation for graphic languages \cite{Rozenberg:HGG:97}.  A graph grammar consists of a set of rules and each rule can be applied if there are any sub-graphs that match.  This idea is derived from the concept of a context-free grammar (CFG), except that a CFG deals with strings while a GG is associated with graphs.  Graph grammars can be used to specify a software model and they provide a natural way of generating instances \cite{hoff:mmgg:10}.  Alternatively, a metamodel graphically describes models and can be considered as a graph in the sense of graph grammars.  Therefore, the natural derivation process of a graph grammar can be adapted for generating metamodel instances.  Thus, the techniques that have been developed in this area to generate sample models are also included in our literature review.

\subsubsection{4. SAT/SMT based Approaches}
SAT/ Satisfiability Modulo Theories (SMT) solvers are adapted to reason about the consistency of a model.  Research work has demonstrated that models can be encoded as boolean logic formulae, and the resulting formulae are transformed into Conjunctive Normal Form (CNF) via a standard transformation, and the final CNFs are solved by a SAT solver \cite{tse:68:opc,dan:asrl:00}.  Each successful assignment for the boolean logic formulae is thus translated back into the problem domain.  The major difference between SAT and SMT solvers is that SMT solvers are supported by its rich background theories such as linear integer arithmetic theory, while SAT solvers only use propositional logic \cite{bar:smtlib:10}.  Using SAT/SMT solvers, a model's properties can be verified within a limited search space by finding an instance.  Since a metamodel is a model that describes another model, the research work that employs SAT/SMT solvers to verify model properties is a highly relevant area.  Within this area, a model is translated into SAT or SMT instances which can be solved by external SAT/SMT solvers, and the solutions can be interpreted back to instances of UML class diagram.

We have already listed the main areas that are related to model instance generation techniques.  We expanded our time period for searching conference proceedings to span the lifetime of a conference series.  We have also included other areas, such as XML instance generation, since a metamodel is represented as a format of XML Metadata Interchange (XMI).  In the following paragraphs, we focus our discussion on algorithms and theoretical frameworks that are used for automatic model instance generation in those areas.

\subsection{RQ2. Within those areas, What algorithms and theoretical frameworks are used for metamodel instance generation?}
In this subsection, we group metamodel instance generation techniques into 6 main areas.  We examine the papers that appear in each area, and discuss the capabilities of their algorithms and theoretical frameworks.

\subsubsection{1. Compiler testing}
In the area of compiler testing,  \citeasnoun{bouj:ctcg:97} present a survey about existing compiler testing techniques.  Most of the techniques in their survey use context free grammars (CFG) to generate test strings.  For example, a sentence generation algorithm generates sentences from a CFG \cite{purdom:bit72}.  In order to generate semantically correct programs,  \citeasnoun{harm:tag:00} propose a framework that generates semantically correct programs using an attributed grammar.  However, this framework only works for small scale of grammars and no evidence indicates that it can be applied to general case.  Much research on grammar testing has already shown a way of direct test case generation from a given language grammar \cite{laemmel:fase01,laemmel:tcs06}.  However, a metamodel captures an abstract syntax of a language which is different from a CFG.  Furthermore, a metamodel is a graph while a grammar is a tree-like structure.  Therefore, the techniques used in testing a compiler can not be directly applied to metamodel instance generation.  

\citeasnoun{alan:arbc:03} propose two algorithms, one to transform a metamodel to a context-free grammar, and the other one to derive a metamodel from a context-free grammar.  Their algorithm for the derivation of a grammar from a metamodel is limited with respect to the structure of the metamodel, as it can only deal with composite association in a metamodel.  By using their algorithm to transform a metamodel into a grammar, it is possible to apply an existing string generation algorithm to get sample strings from the grammar, and then translate the sample strings back into instances of the metamodel.  However, if we define a metamodel that preserves properties of a UML class diagram, it is not surprising that their algorithms cannot deal with multiplicities of an association.  Furthermore, translating generated sample strings from a grammar to instances of the metamodel would be difficult since a metamodel (graph) and a grammar (tree) have different structures. 

\subsubsection{2. Model transformation testing}
In the area of model transformation testing, some of the techniques used for automatic model instance generation have been developed.  \citeasnoun{bro:06:mtg} propose an algorithm to automatically generate a set of model instances for testing a model transformation.  In this algorithm, the authors use two concepts: model fragments and object fragments.  To form an object fragment, they employ the classic category-partition testing technique \cite{os:88:cpmft}.  They partition the metamodel into different equivalence classes by using coverage criteria for testing the UML class diagram \cite{andrews03:stvr}.  For example, an integer type attribute $P$ of a class $C$ is partitioned into three categories \{$P=0$, $P=1$, $P>1$\}.  Each element in the set is marked as an object fragment, and an object fragment specifies a partial instance of the class $C$.  A model fragment then consists of a list of such object fragments.  The algorithm then takes in a set of model fragments and outputs a set of models that conform to the metamodel.

The algorithm of \citeasnoun{bro:06:mtg} iterates over the uncovered model fragments and keeps generating model instances until all the model fragments are covered.  Furthermore, this algorithm provides some strategies for testers to meet specific testing needs.  For example, it provides a strategy that reuses generated model instances as much as possible to achieve the smallest number of instances to cover most of the model fragments.  However, this algorithm has two main limitations. One is that it supposes that all the model fragments are manually provided by testers.  The other is it does not take Object Constraint Language (OCL) constraints into account \cite{warmer99:ocl}.  Therefore, the constraints defined in the metamodel are overlooked and, as a result, the generated model instances do not satisfy the constraints.  

\citeasnoun{fle:vme:04} propose metamodel coverage criteria and two algorithms that can be used to automatically generate metamodel instances.  The coverage criteria will be discussed in Section 5.3.  The two algorithms they propose are the systematic approach and the bacteriologic approach \cite{bro:06:mtg}.  We discussed the first algorithm in the previous paragraph.  The second algorithm they proposed is a bacteriologic algorithm which is based on genetic algorithms.  This algorithm consists of 4 steps.  Before executing the algorithm an initial set of test models have to be provided.  In the first step of the algorithm, the test models are ranked using a fitness function which estimates the contribution of the coverage they can make.  In step 2, the test models which make the most contribution to the coverage are memorised and added to the solution set.  The test models which can not make any contribution to the coverage are removed in step 3.  In the final step, a mutation operator is applied to the best test models in order to create new test models.  The algorithm keeps iterating until all coverage items are covered by test models.  \citeasnoun{baud:atc:02} implemented this algorithm.  However, the limitation of this algorithm is it requires an initial set of test models.  Therefore, to compute such a set of test models is another problem of instance generation.

\subsubsection{3. Graph Grammar}
We have also studied research retrieved from the area of graph grammars.  \citeasnoun{ehri:gimm:09} extend the work by \citeasnoun{bard:deLara:04}, and propose an instance-generating graph grammar for creating metamodel instances.  In their approach, they use the concept of layered graph grammars to order rule applications.  Each rule is applied when a specific metamodel pattern is found and the corresponding application conditions are satisfied.  Three layers are defined for the rules.  The rules in layer 1 are used for creating instances of each non-abstract class in the metamodel, and layer 2 consists of a set of rules that deal with three metamodel patterns.  These patterns are those association relationships defined in the metamodel with multiplicity of one.  The rules in layer 3 are defined for taking care of associations with multiplicity having a bound of zero to $n$.   They illustrated this approach by applying these rules on an example.  

However, in the work of \citeasnoun{ehri:gimm:09} composition associations are not considered.  This was later tackled by \citeasnoun{hoff:11:gcm}.  This work uses graph grammars called adaptive star grammars \cite{drew:asg:06} to show how a metamodel (represented as a class diagram) can be translated into this grammar, and such a graph grammar can be used to automatically generate instances of the metamodel.  In this approach, a set of adaptive star grammar rules are introduced to deal with relationships defined in the metamodel, such as unique, non-unique and composition associations.  However, OCL constraints defined over the class diagram can not be handled by this approach.

To deal with OCL constraints, \citeasnoun{win:tro:08} present a method in which OCL constraints can be translated into graph constraints.  In their approach, they show how a list of OCL constraints can be translated into equivalent graph constraints.  These OCL constraints are restricted to equality, size and attribute operations for navigation expressions, called restricted OCL constraints.  They suggest two ways of generating a valid model instance that satisfies the OCL constraints.  One way is to check the translated constraints once after an instance is generated.  The other way is taking the constraints into consideration during the instance model generation process.  Both ways have their advantages and disadvantages.  The former may lead to a large number of invalid model instances are generated more quickly, while the later produces only those model instances which satisfy the constraints and are generated more slowly.

\subsubsection{4. SAT/SMT based Approaches}
{\sf Alloy}, a model finder, is a mature tool that can be used for finding instances of a model and counter examples in a finite search scope \cite{alloy4,dan:02:alloy}.  {\sf Alloy}, unlike other model-finding tools such as MACE \cite{mace:03}, uses relational logic to describe a model.  To speed up searching process {\sf Alloy} bounds relational logic, translates the bounded relational logic into a boolean formula, and then it calls an off-shore SAT solver to solve the boolean formula.  The solutions returned by the SAT solver then get translated back into the models \citesec{Torlak:CSS:09} \cite{torlak:06:thedesign}.

Since {\sf Alloy} can be used to generate instances of a model within a bounded search space, research with {\sf Alloy} has been highly active \cite{bord:ates:05,Ana:10:sos,Anas:uml2alloy:07,sen:amgsmtt:09,mcquillan08:stvv}.  \citeasnoun{Anas:uml2alloy:07} focus on transforming UML class diagrams to the {\sf Alloy} language.  In their work, they demonstrate a list of rules which can map a UML class diagram and the limited number of OCL constraints to the {\sf Alloy} language.  \citeasnoun{sen:ocm:08} are inspired by this approach and present a tool called {\sf Cartier} to automatically generate test models for testing model transformations.  This tool can transform a metamodel (in Ecore format) with OCL constraints, model transformation pre-conditions, model fragments \cite{fle:qit:09} and test objectives into the {\sf Alloy} language.  They also present some strategies based on a partition techniques to guide the tool to generate models \cite{sen:amgsmtt:09}.  These strategies are written in {\sf Alloy} predicates, and combined with models are solved by SAT solvers.  The solutions returned by SAT solvers are then transformed by the {\sf Cartier} tool back to the instances of the input metamodel.

Another related approach is by \citeasnoun{mcquillan08:stvv}.  They propose a metric metamodel for the measurement of object-oriented software.  In their work, they firstly transform the metrics metamodel along with the OCL constraints into an {\sf Alloy} specification, and then use {\sf Alloy} to generate possible instances.  To transform the {\sf Alloy} generated instances back to the metamodel instances, they developed a tool called Reflective Instantiator.  This tool reads an {\sf Alloy} specification file and creates an instantiation of the Java implementation of the metamodel.

Both the approaches of \citeasnoun{sen:ocm:08} and \citeasnoun{mcquillan08:stvv} require the pre-processing of a metamodel to {\sf Alloy}, followed by a process that transforms from the {\sf Alloy} output back to the instances of the metamodel.  This increases the complexity of the model instance generation process, especially if one wants to visualise the instances, for example, if one wants to transform the instances to UML class diagrams.  Furthermore, both tools utilise  {\sf Alloy}'s APIs to invoke the model generation process, thus both tools are not easy to maintain since they are highly depended on {\sf Alloy}.

Many SMT-Solves are well developed and supported by multiple theories, for example lists, sets etc \cite{bru:mss:08,brutt:opensmt:10,demou:z3:08}.  Compared to SAT solvers, SMT-Solvers have a greater advantage because of multiple theories defined in the Satisfiability Modulo Theories Library (SMT-Lib) \cite{bar:smtlib:10}.  Unlike SAT Solvers which take in a Conjunctive Normal Form (CNF) as their input, SMT-Solvers take in a formula in a more natural way according to a specific theory.  A typical example is solving a linear integer arithmetic equation.  Thus, we have also studied approaches that take an advantage of SMT-Solver to find instances of a model \cite{soe:10:vumubs,Soeken:11:EOD,eth:11:Formula}.

\citeasnoun{soe:10:vumubs}, encode a UML class diagram as a set of operations on bit-vectors which can be solved by SMT solvers using bit-vector theory.  A successful assignment for each bit-vector can be interpreted as an instance of the UML class diagram.  \citeasnoun{Soeken:11:EOD} propose an approach to encode a subset of OCL constraints as bit-vectors, and provide a list of corresponding mappings between OCL collection data types (Set, Bag, Sequence) and bit-vector operations.  Furthermore,  to allocate an appropriate bound for each entity defined in the model is essential for bounded model checking and verification \cite{Soe:11:TAD}.  \citeasnoun{Soe:11:TAD} propose a linear integer arithmetic approach.  This approach considers different kinds of associations defined in the model, translates defined multiplicities in each association into linear integer arithmetic equations, and uses a SMT solver to solve the equations.  The bound for each entity can therefore be determined by solved equations.  However, this approach cannot deal with OCL invariants that specify the number of instances to be created.

\citeasnoun{eth:11:Formula} propose a framework called Formula, which is a MOF-like framework that can capture the metamodelling's abstraction via a graph-like language.  This framework consists of three components: a model store to specify models and metamodels, a list of operations to edit models and metamodels, and a meta-interpreter which can promote model-level elements to meta-level elements.  To be able to verify properties of operations on models and metamodels, the generated constraints are solved by a SMT-Solver \cite{demou:z3:08}.

We have also considered other work that takes an advantage of SMT-Solvers to complement {\sf Alloy} techniques.  \citeasnoun{elg:rrv:11} present work that translates an {\sf Alloy} specification into SMT-instances to prove {\sf Alloy} assertions.  This work also takes an advantage of linear integer arithmetic which is supported by a SMT-solver to perform unbounded integer arithmetic operations.

\subsubsection{5. Constraint Programming Approach (CSP)}
\citeasnoun{cab:vuc:08} propose a procedure that can transform a UML class diagram with OCL constraints into a Constraint Satisfaction Problem (CSP) according to a set of rules.  The CSP is described by using the syntax provided by the ECL$^{i}$PS$^{e}$ Constraint Programming System \cite{apt:clp:07}.  The CSP itself can be divided into two subproblems.  The first subproblem is to determine a bound for each class and association (variables), and the second is to assign a value to each variable.  If CSP has a solution (instance), the user can conclude that a model satisfies the properties.  \citeasnoun{cab:09:vuoc} extend their work to OCL operation contracts by using a similar translation process to the one used in their earlier work \cite{cab:vuc:08}.  Their work is supported by a tool called UMLtoCSP, and a tool called EMFtoCSP which extends UMLtoCSP to deal with EMF metamodels \cite{cab:uml2csp:07,gon:emf2csp:12}.  Both tools are designed for bounded verification and they do not provide an instance enumeration mechanism like {\sf Alloy}.

\subsubsection{6. XML}
A MOF metamodel is defined in the format of XML Metadata Interchange (XMI) which is an Object Management Group (OMG) standard for exchanging metadata information via the Extensible Markup Language (XML) \citesec{omg:mof:10}.  Therefore, the work done in this area is also related.  \citeasnoun{bert:atd:07} propose an approach called XML-based Partition Testing (XPT) to automatically generate XML instances from a XML Schema.  This approach uses the idea of a classic Category Partition method to derive a set of XML instances from a preprocessed XML Schema \cite{os:88:cpmft}.  

The XPT methodology consists of five steps and is partially implemented in a proof-of-concept tool called ``Testing by Automatically generated XML Instances (TAXI)" \cite{bert:sgx:06}.  In the first step, a XML Schema is preprocessed and some of the elements in the XML Schema are rewritten in order to facilitate the later steps.  Next, an occurrence analysis is adopted to manipulate the boundary values for the occurrence of each element in the XML Schema.  The third step is to assign the value to each element.  These values are either randomly generated or selected from a database.  A set of intermediate instances are then generated by combining the occurrences of each element in the fourth step.  Finally, the actual XML instances are generated from the intermediate instances.  However, this approach does not handle the associations defined in the model and furthermore it does not consider any OCL constraints defined in the model.  

\subsubsection{7. Others}
\citeasnoun{gogo:vuomu:05} propose an approach that uses the language ASSL.  ASSL is an extension language of the USE specification language \cite{gogolla07:scp}.  It provides a list of commands for testers to generate snapshots for finding defects in a UML class diagram.  A snapshot is an object diagram that represents the system states at any time with objects, attribute values and links.  The snapshots can be created by an ASSL procedure, and moreover the snapshots can be created with consideration of the presence of OCL invariants.  With their approach, it is possible to generate a number of snapshots to validate the design at an early stage.  However, one must know how to write an efficient ASSL procedure in order to generate appropriate snapshots.  

\citeasnoun{mou:urg:09} use an approach that is based on the Boltzmann method \cite{duch:bsr:04} to generate metamodel instances.  In their approach, a metamodel is first transformed into a Boltzmann tree specification and the final instances are derived from the generated trees.  The main purpose of this approach is to make sure that the instance generation process has no bias, and therefore the generated instances are uniform.  Another advantage of this approach is that the Boltzmann method has a linear property, which means that the instance generating process is linear with respect to the size of the generated instances.  However, this approach does not take into account any constraints defined in the metamodel.  Hence, instances that do not satisfy constraints can be generated.

\citeasnoun{lukman:agmtfmd:10} propose a traversal algorithm for metamodels.  In their algorithm, the root element of a metamodel is first identified.  A root element in a metamodel is a container class that can be used to form a composition relationship with other classes.  This algorithm traverses each connected class by association, generalisation and composition relationships defined in the metamodel, and marks the paths that have been visited to prevent an infinite iteration of each class.  Although their work does not describe how to generate a model instance, by applying this algorithm it is possible to generate an instance from each class that has been traversed.  However, traversing a metamodel is not sufficient to produce valid instance from the metamodel.  For example, a valid instance must meet the multiplicities of an association, which requires the algorithm to be aware of the links between objects.

\subsection{RQ3. What criteria are used for selecting model instances?}
\citeasnoun{fle:vme:04} propose metamodel coverage criteria.  This originates from existing coverage criteria that have been adapted from UML class diagrams.  Since a MOF metamodel is similar to a UML class diagram, they reuse existing UML coverage criteria for metamodels.  These coverage criteria were firstly proposed by \citeasnoun{andrews03:stvr} and are defined for a UML class diagram using three criteria:

\begin{compactenum}
\item[1.] Association-end multiplicities (AEM): Each representative link in a class diagram must be covered.
\item[2.] Class attribute (CA): In each instantiated class, the set of representative attribute value combinations in each instance of a class must be covered.
\item[3.] Generalisation (GN): Every specialisation defined in a generalisation relationship of a class diagram must be covered.
\end{compactenum}

Two of the criteria listed above (AEM and CA) use representative values that are based on partition testing techniques to express their meanings.  The representative values can be selected by applying knowledge-based or default partitioning approaches.  For example, an integer attribute can be partitioned into three values which are greater, less then and equal to zero.  For knowledge-based partitioning, the values can be provided by testers or determined by the specification of the model, e.g. by examining the constraints defined over the model.  For default-partitioning, the representative value can be selected by partitioning an attribute into minimum, non-boundary and maximum values.  \citeasnoun{fle:vme:04} adapt these two criteria to achieve metamodel coverage and define their test criterion.  Later this criterion was implemented in the work \citeasnoun{bro:06:mtg}.

To continue the work in \citeasnoun{bro:06:mtg}, \citeasnoun{fle:qit:09} proposed an approach to select model fragments, a list of test criteria and a tool developed for automatically generating model fragments.  They also designed a generic test criteria metamodel, and the generated model fragments from the tool must conform to this metamodel.  To evaluate whether a set of test models satisfy the criteria defined by the test criteria metamodel, it uses a tool called {\sf MMCC} which involves four steps.  In the first and second step, the tool automatically generates a set of model fragments according to the test criteria.  Then in step 3 these generated model fragments are checked against the test models.  The final step shows uncovered model fragments, if it there are any, and requires a manual analysis of the reasons why those uncovered model fragments are left.  The test criteria defined in their approach are defined using OCL constraints over the test criteria metamodel.  They propose 6 criteria are shown in Table \ref{tab:criteria}:
\\
\begin{table}
\begin{tabular}{l | p{5cm} | p{3cm} }
	\toprule
	Name of the criteria & Description  & Type\\ \hline
	1. AllRanges & All ranges of a property of a class to be covered. & NA \\ \hline
	2. AllPartitions & All ranges of a property of a class to be covered in one test model. & NA \\ \hline
	3. OneRangeCombination & Every range of each property of a class have to be covered at least once. & Object Fragment\\ \hline
	4. AllRangeCombination & One object fragment must contain all possible combinations of ranges of a property of a class. & Object Fragment \\ \hline
	5. OneMFPerClass & A single model fragment must contain all possible combinations of ranges for a class. & Model Fragment\\ \hline
	6. OneMFPerCombination & Every model fragment only contains one possible combination of ranges for a single class. & Model Fragment \\ \bottomrule
	
\end{tabular}
\caption{6 test criteria defined in Fleurey et al. (2009)}
\label{tab:criteria}
\end{table}
\\
Among the 6 criteria listed in Table \ref{tab:criteria}, those criteria marked with the type ``object fragment" can be combined with those marked as a ``model fragment" to form the test criteria.  For example, the combination of criteria 3 and 5 results in one model fragment that contains all possible combinations of ranges of properties of a class.  However, the criteria defined in their approach produce quite a high number of object fragments since they use a Cartesian product on ranges.  This may lead to a combinatorial explosion if the input metamodel has a large number of properties defined in each class. 

\subsection{RQ4. What tools exist to produce model instances?}
A number of tools have been developed for producing or assisting automatic instance generation.  The tools that support model instance generation can be mainly grouped into two categories: independent and dependent.  An independent tool can be run without the support of any other tools or platforms, while a dependent tool requires some support.  The tools are given in Table \ref{tab:tools} with a column to indicate their category.  
\\
\begin{table}
\begin{tabular}{ p{8cm} | l }
\toprule
Name of the tool and Reference  & Category \\ \hline
Alloy \cite{dan:02:alloy}, \citesec{jackson:aidl:98,dan:asrl:00,torlak:krm:07} & \\
Formula \cite{eth:11:Formula} & \\
OMOGEN \cite{bro:06:mtg}, & Independent \\
MMCC \cite{fle:qit:09},  & \\ 
Reflective instantiator \cite{mcquillan08:stvv}, & \\
USE \cite{gogo:vuomu:05} & \\ \hline
UML2Alloy \cite{bord:ates:05,Anas:uml2alloy:07} & Depends on Alloy\\
Cartier \cite{sen:amgsmtt:09},  & Depends on Alloy\\
USE model Validator plugin \cite{kuh:11:usekodkod}, & Depends on kodkod\\
UMLtoCSP, EMFtoCSP \cite{cab:uml2csp:07,gon:emf2csp:12}, & Depends on ECL $^{i}$PS$^{e}$ \\
A generator as an Eclipse Plugin \cite{mou:urg:09} & Depends on Eclipse\\
\bottomrule
\end{tabular}
\caption{Tools that support or assist automatic model instance generation}
\label{tab:tools}
\end{table}
\\
{\sf Alloy} is a popular tool used by many modelers to find instances or counter examples of a model.  It was introduced in \citeasnoun{jackson:alcoa:00} and it uses a relational logic to capture the semantics of first-order logic, and thus a model is described in a relational logic \cite{jackson:aidl:98}.  The later version is an improved version that allows modelers to specify quantifiers over the relational logic \cite{dan:02:alloy}.  The latest version of {\sf Alloy} employs a new engine (kodkod) that represents a model as a bounded relational logic, then translates it into a boolean formula that is represented by compact boolean circuits \cite{torlak:krm:07} \cite{Torlak:CSS:09}, and finally solves the formula by using off-shore SAT-solvers \cite{lintao:zchaff:01,MiniSatSystemDesc,dan:sat4j:10}.  With the new engine, {\sf Alloy} is able to deal with larger models and to find instances or counter examples in seconds.  However, one of the main problems with using {\sf Alloy} to find model instances is that it requires a full mapping from the MOF metamodels to an {\sf Alloy} specification.  Since they are in different formats, such mappings may not be always straightforward.  Though much work has been done to deal with the mapping from metamodels to an {\sf Alloy} specification, it still requires considerable work to handle the differences between them \cite{bord:ates:05,Anas:uml2alloy:07}.

{\sf USE} is a tool that can be used to build a variety of UML diagrams such as UML class diagrams, sequence diagrams, etc \cite{gogolla07:scp}.  It also provides the user with a way of creating and checking OCL constraints defined over the model, and fully supports OCL 2.0 \cite{omg:ocl:20}.  The latest research for {\sf USE} involves integrating it with {\sf Alloy}'s core engine, kodkod \cite{kuh:11:usekodkod}.  The resulting integration with kodkod yields a {\sf USE} Model Validator plugin.  This plugin provides users with an interface that allows a user to specify a lower and upper bound for each class defined in the UML model.  Since this approach is fully depended on the performance of kodkod, it thus requires a translation from UML class diagrams and OCL constraints into relational logic, which complicates the instance generation process.

\citeasnoun{bro:06:mtg} implement their algorithm into a tool called {\sf OMOGEN}.  This is a prototype tool and it can be used to generate model instances based on the input of a set of model fragments.  However, to generate the model fragments a tool called {\sf MMCC} is needed \cite{fle:qit:09}.  This tool is implemented using Eclipse Modeling framework (EMF), and it generates partitions from a source metamodel and model fragments according to particular test criterion which were described in Section 4.3.

The {\sf reflective instantiator} described in \cite{mcquillan08:stvv}, parses generated instances from {\sf Alloy} into instances of the Java implementation of the metamodel.  Furthermore, this tool generates a code implementation of a metamodel. 

\citeasnoun{gogo:vuomu:05} extend {\sf USE} features by defining a new language to generate snapshots of the models.  These snapshots are treated as test cases and can be validated within {\sf USE}.  Their approach allows OCL invariants to be dynamically loaded and certified during the execution.  However, it does not provide any test criteria during the generation of snapshots and the test criteria have to be manually programmed.  

The {\sf Cartier} Tool \cite{sen:amgsmtt:09} depends on {\sf Alloy}.  It invokes {\sf Alloy} APIs to launch the SAT solver to generate model instances.  However, {\sf Cartier} does not automatically transform the OCL constraints of the input metamodel into {\sf Alloy} facts.  Furthermore, {\sf Cartier} requires that a set of model fragments is generated before generating model instances.  These model fragments are generated by the tool introduced in \citeasnoun{fle:qit:09}.  Therefore, {\sf Cartier} heavily depends on other tools and this makes the model generation process quite complicated.  

The Boltzmann method is implemented in a generator as an Eclipse Plugin \cite{mou:urg:09}.  This generator produces a valid model by identifying properties, generalisation and references in the metamodel.  This tool is very effective for generating a large uniform sample of models.  However, this tool can not generate models that will satisfy the constraints defined with respect to the metamodel.

\subsection{Summary}
Most of the research done on instance (model) generation techniques comes from three areas: model transformation testing, graph grammars and SAT/SMT based approaches.  In the area of model transformation testing, algorithms, frameworks and tools have been developed for generating or assisting metamodel instance generation \cite{fle:vme:04,bro:06:mtg,fle:qit:09,lami:tat:07,Sen:MMS:06}.  However, none of them can handle OCL constraints with respect to the metamodels.  

To be able to handle OCL constraints, SAT/SMT seems to be more suitable for adapting because OCL constraints can be more naturally expressed as boolean formula than other approaches.  Among these approaches, {\sf Alloy} seems the most mature tool to be adapted and many researchers have carried out work using {\sf Alloy} \cite{sen:ocm:08,sen:amgsmtt:09,mcquillan08:stvv}.  However, {\sf Alloy} and the metamodelling technique employ different languages, and therefore a mapping between them is inevitable.  Unfortunately, such a mapping is not always straightforward, for example a one-to-one binary association requires users to write additional facts.  Though some of the research work has already been conducted, such a mapping can only increase the complexity of the generation process \cite{bord:ates:05,Anas:uml2alloy:07}.  Furthermore, using {\sf Alloy} to generate instances to meet the specific coverage criteria in the first place is difficult, since {\sf Alloy} enumerates all ``possible'' solutions.  Although this can be achieved by defining facts in {\sf Alloy}, this may slow down the process of finding instances, and it requires that a user has a good knowledge of writing efficient facts to filter out unnecessary solutions.

Another area which has been exploited to overcome the disadvantage of metamodelling is the area of graph grammars.  Graph grammars offer a natural way to describe the derivation process and so have an advantage for generating metamodel instances.  However, before applying grammar rules, a map from metamodels to graphs is also needed.  Furthermore, parsing a graph is expensive in terms of an algorithm's complexity because a graph matching is not always deterministic, as a rule may match several sub graphs \cite{ehrig:fag:06}.  Another issue is that the current techniques that exist in the graph grammar area can only deal with a limited number of OCL constraints.  These OCL constraints must be manually transformed into graph constraints which causes a problem when one has to deal with a large number OCL constraints \cite{win:tro:08}. 

By analysing and understanding these papers we collected, we have identified that the knowledge gap between metamodel instance generation and metamodelling techniques is that there is no direct way of generating metamodel instances from a metamodel, as all techniques either require a pre-process or extra knowledge about an intermediate language to achieve instance generation.  Thus, it requires tools or mappings to naturally support an instance generation process.  It would be a valuable research direction to discover how such a framework or tools could be developed, and could be directly applied to not only generate model instances, but also to meet the constraints defined over the metamodels.  Furthermore, it will also be useful to construct such an extensible framework or tool so that the generated instances can meet specific coverage criteria.  

\section{Conclusion}
In this paper, we conducted a systematic literature review on metamodel instance generation, and the results are also shown in the paper.  The areas that are related to metamodel instance generation techniques were identified and papers from those areas were discussed in detail with a view to answering our research questions.  A knowledge gap in the area was also identified as a potential future research direction.  The papers collected in this literature review  cover the areas of metamodel instance generation techniques but we believe that it identifies most of the core papers and the work done in this paper can be easily extended if any core papers appear in the coming years.


\bibliographystyle{agsm}

\end{document}